\documentclass[prc,aps]{revtex4}  \usepackage{graphicx} 
\begin{document} 
\pagestyle{plain} 
\draft
\title{Proton skins, neutron skins, and proton radii of mirror nuclei.                                                
}
\author{            
Francesca Sammarruca      }                                                                                
\affiliation{ Physics Department, University of Idaho, Moscow, ID 83844-0903, U.S.A. 
}
\date{\today} 
\begin{abstract}
We present predictions for proton skins based on isospin-asymmetric equations of state derived microscopically from
high-precision chiral few-nucleon interactions. Moreover,
we investigate the relation between the neutron skin of a nucleus and the difference between
the proton radii of the corresponding mirror nuclei.                    
\end{abstract}
\maketitle

\section{Introduction} 
\label{Intro} 
It is well known that the available information on neutron radii and neutron skins is scarce and carry considerable uncertainty, see, for instance, Ref.~\cite{FS3} and references therein for a summary of empirical constraints,
particularly on the skin of $^{208}$Pb, obtained from a variety of measurements. 
Although future experiments~\cite{Jlab,CREX} are anticipated which should provide reliable information on the 
weak charge density in $^{208}$Pb and  $^{48}$Ca, the identification of other ``observables" whose knowledge may give complementary information on neutron skins would be most welcome.        

Naturally, the possibility of obtaining reliable values for neutron or 
proton skins is hindered by similar limitations, as both proton and neutron radii must be known to extract 
either skin. And while charge densities, particularly for stable isotopes, have been measured with great
accuracy, the same cannot be said for the weak charge density. 

An issue of current interest is whether information on the neutron skin can be obtained through the  
knowledge of proton radii alone, specifically those of mirror nuclei. In particular, the difference between the charge radii of mirror nuclei in relation to the slope of the symmetry energy, and, in turn, to the neutron skin, was nvestigated in Ref.~\cite{AB17}.  As done in the past by the same 
author~\cite{AB00}, correlations between neutron skins and the slope of the symmetry energy       
are deduced using large sets of phenomenological interactions, such as the 
numerous parametrizations of the Skyrme interactions. In Ref.~\cite{AB17}, using 
similar methods and 48 Skyrme functionals, a proportionality relation was
found between the difference in the charge radii of mirror nuclei  
and the slope of the symmetry energy.  This was echoed in Ref.~\cite{YP} using 
a set of relativistic energy density functionals.

Although phenomenological analyses are a useful exploratory tool to gain some preliminary 
insight into sensitivities and interdependences among nuclear properties, 
only through microscopic predictions can we understand a result in terms of 
the physical input.
The purpose of this work is twofold. First, we present proton skin predictions and observe general 
patterns within isotopic chains, comparing with data when available. 
Second, we wish to explore, from 
the microscopic point of view in contrast to the phenomenological one, the relation between 
the neutron skin of a nucleus, on the one hand, and the difference between the proton radii of the 
mirror pair with the same mass, on the other. To avoid confusion, we underline that this analysis will not 
be done by varying parameters in a family of models (not an option 
consistent with the microscopic approach). Instead,
we shall investigate our predicted relation between the quantities defined above for 
a variety of (realistic) mirror pairs.
We will pay particular attention to proton radii of mirror nuclei specifically in the mass range
$A \approx 48-54$. 
 At this time, the determination of proton radii of neutron-deficient isotopes such as, for instance, the ``mirror"
of $^{54}_{26}$Fe is  an enormous experimental challenge, which may be met in the future at radioactive beam facilities. 

Our predictions are based on microscopic high-precision nuclear interactions derived from chiral Effective Field Theory (EFT)~\cite{ME11}. 
In this way, we hope to provide useful  microscopic input to be taken into account in future analyses.

This paper is organized as follows. In the next section, we give a short review of the 
theoretical tools and the calculation of the neutron and proton skins. 
We then proceed to proton skin predictions (Sect.~\ref{res}) and, more specifically, those of some
mirror pairs in selected mass ranges (Sect.~\ref{mirr}).  A brief 
summary and our conclusions are contained in Sect.~\ref{Concl}. 

\section{Brief review of the theoretical input } 
\label{theor} 

\subsection{The few-nucleon forces} 
\label{few} 

In recent years, chiral EFT has evolved into the authoritative approach   
to construct nuclear two- and many-body forces in a systematic and essentially model-independent 
manner~\cite{ME11,EHM09}.   
Nucleon-nucleon (NN) potentials are available from leading order (LO, zeroth order) to
N$^3$LO (fourth order)~\cite{ME11,EHM09,NLO,EM03,EGM05}, with the latter reproducing NN data at the high precision 
level. More recently, NN chiral potentials at N$^4$LO have also been developed~\cite{n4lo,EKM15}. 

A large number of applications of chiral NN potentials (usually up to N$^3$LO) together with chiral three-nucleon
forces (3NF) (generally just at N$^2$LO) have been conducted. A fairly extensive, although not exhaustive list
is given in Refs.~\cite{NCS,cc1,cc2,cc3,SRG,Stro16,ab1,ab2,ab3,ab4,Cor+,ab5,Soma,ab5a,ab6,ab7,ab8,ab9,ab10,CEA,hag+,kth,obo}.

We apply  the microscopic equations of state (EoS) of symmetric nuclear matter and the ones of pure neutron matter as derived in Ref.~\cite{obo}. 
 The derivation is based on high-precision chiral NN potentials                            
at next-to-next-to-next-to-leading order (N$^3$LO) of chiral perturbation theory~\cite{EM03,ME11}.
The leading 3NF, which is treated as an effective density-dependent           
force~\cite{HKW}, is included.

\subsection{Additional tools} 
\label{rev} 
This section provides a very brief summary of previously developed 
tools to obtain 
nuclear properties from the infinite-matter EoS~\cite{AS03}. Within the spirit of a liquid 
droplet model, the energy of a          
 nucleus is written in terms of a volume, a surface, and a Coulomb term as 
\begin{equation}
E(Z,A) = \int d^3 r~ e(\rho,\alpha)\rho(r) + 
\int d^3 r f_0|\nabla \rho|^2 +        
\frac{e^2}{4 \pi \epsilon_0}(4 \pi)^2 
\int _0^{\infty} dr' r' \rho_p(r')       
\int _0^{r'} dr r^2 \rho_p(r) \; .  
\label{drop} 
\end{equation} 
In the above equation, 
$\rho$ is the total nucleon density, given by $\rho_n +\rho_p$, with 
$\rho_n$ and $\rho_p$ the neutron and proton densities, respectively. 
$\alpha$ is the neutron asymmetry
parameter, $\alpha=\rho_I/\rho$,                                                     
where the isovector density $\rho_I$ is given by $(\rho_n -\rho_p)$.                                    
$e(\rho,\alpha)$ is the energy per particle in 
isospin-asymmetric nuclear matter, written as 
\begin{equation}
e(\rho,\alpha) =                                  
e(\rho,0) + e_{sym}(\rho)\alpha^2 \; ,             
\label{eee}  
\end{equation}
with $e_{sym}(\rho)$ the symmetry energy. 
The density functions for protons and neutrons are obtained by minimizing the value
of the energy, Eq.~(\ref{drop}), with respect to the paramaters of Thomas-Fermi distributions, 
\begin{equation}
 \rho_i(r) = \frac{\rho_0}{1 + e^{(r-a_i)/c_i}} \; , 
\label{TF}  
\end{equation}
with $i=n,p$. The radius and the diffuseness, $a_i$ and $c_i$, respectively, are extracted by minimization 
of the energy while $\rho_0$ is obtained by normalizing the proton(neutron) distribution to $Z$($N$). 
The neutron and proton skins are defined in the usual way, 
\begin{equation}
 S_n = R_n - R_p \; ,                                 
\label{rnrp}   
\end{equation}
and 
\begin{equation}
 S_p = R_p - R_n \; ,                                 
\label{rnrp}   
\end{equation}
respectively, 
where $R_n$ and $R_p$ are the $r.m.s.$ radii of the neutron and proton density distributions, 
\begin{equation}
 R_i = \Big ( \frac{4 \pi}{T} \int_0 ^{\infty} \rho_i(r) r^4 \; dr \Big )^{1/2} \; , 
\label{rms}   
\end{equation}
and $T$= $N$ or $Z$. 
We stress that the above method has the advantage of allowing for a         
very direct connection between the EoS and the properties of finite nuclei. It was used in Ref.~\cite{AS03} in 
conjunction with meson-theoretic potentials and found to yield realistic predictions for binding 
energies and charge radii. 
The constant $f_0$ in the surface term is typically obtained from fits to $\beta$-stable nuclei and determined to be about 60-70 MeV fm$^5$~\cite{Oya2010}. How this uncertainty impacts the corresponding predictions was discussed in Ref.~\cite{FS3} and will be taken into account in the present calculations.

\section{Predictions for proton skins} 
\label{res} 

In Table~\ref{tab1}, we display proton skin predictions for some isotopic chains.
The EoS used for these predictions is based upon N$^3$LO two-nucleon
forces (2NF) plus the leading 3NF.
The estimated theoretical errors include uncertainties due to variations of the cutoff in the range 450-500 MeV
as well as an error (added in quadrature) to account for the uncertainty
originating from the method we use to calculate the skins~\cite{FS3}. 
The latter error is in the order of $\pm$ 0.01 fm, but varies with the size of the skin.         

As a general feature, we observe that the proton skins can be quite large. In fact, the {\it neutron} skins of                    
the corresponding (neutron-rich) mirror nuclei are smaller.
This fact is demonstrated in Table~\ref{tab2}, where 
we show, for the most neutron-deficient isotope in each chain, 
the proton skin together with the neutron skin of the corresponding mirror nucleus.        

Some data on proton skins can be found in Refs.~\cite{Suz+95,Suz+98,Oz+02,AW95}. In
Ref.~\cite{Suz+98}, the existence of neutron and proton skins in 
$\beta$-unstable neutron- or proton-rich Na and Mg isotopes is discussed based 
on measurements of the interaction cross sections of these isotopes incident on a carbon target around 950$A$ MeV.
In Ref.~\cite{Oz+02}, proton skin thickness for 
isotopes $^{32-40}$Ar were deduced from the interaction cross sections of
$^{31-40}$Ar and $^{31-37}$Cl on carbon targets. The obtained matter radii were combined with measured charge radii for Argon isotopes to obtain skin thicknesses.                                                                 

In Fig.~1, we show our predictions for the proton skins 
of Argon isotopes in comparison with data deduced from experiments as
described in Ref.~\cite{Oz+02}. 
Keeping in mind the large experimental errors, the trend of the empirical information is described reasonably well by our predictions, where the proton skin decreases essentially monotonically with increasing number of neutrons in a given isotopic chain.

\begin{table}                
\centering
\caption                                                    
{ 
 Proton skins, $S_p$, for $Z$=10, 11, 17, and 18 isotopic chains. 
See text for more details. 
} 
\begin{tabular}{l l c} 
\hline
\hline
Z  \hspace{0.8in} & A \hspace{0.8in} &   $S_p$ (fm) \\ 
\hline     
10 &16 &    0.422 $\pm$ 0.022 \\ 
   &17 &   0.287 $\pm$ 0.014 \\ 
   &18 &   0.186 $\pm$ 0.012 \\ 
   &19 &   0.103 $\pm$ 0.006 \\ 
   &20 &  0.032 $\pm$ 0.006 \\ 
11 &18 &  0.373 $\pm$ 0.020 \\ 
   &19 &  0.260   $\pm$ 0.012 \\ 
   &20 &  0.172   $\pm$ 0.012 \\ 
   &21 &   0.098   $\pm$ 0.006 \\ 
   &22 &  0.034   $\pm$ 0.006 \\ 
17 &31 &  0.180 $\pm$ 0.012 \\ 
   &32 &  0.131 $\pm$ 0.011 \\ 
   &33 &  0.086 $\pm$ 0.008 \\ 
   &34 &  0.045 $\pm$ 0.007 \\ 
18 &29 &  0.439 $\pm$ 0.025 \\ 
   &30 &  0.352 $\pm$ 0.019 \\ 
   &31 &  0.283 $\pm$ 0.014 \\ 
   &32 &  0.225 $\pm$ 0.013 \\ 
   &33 &  0.174 $\pm$ 0.013 \\ 
   &34 &  0.127 $\pm$ 0.012 \\ 
   &35 &  0.085 $\pm$ 0.008 \\ 
   &36 &  0.046 $\pm$ 0.007 \\ 
\hline
\hline
\end{tabular}
\label{tab1}
\end{table}

\begin{figure}[!t]
\vspace*{1cm}
\includegraphics[width=8.5cm]{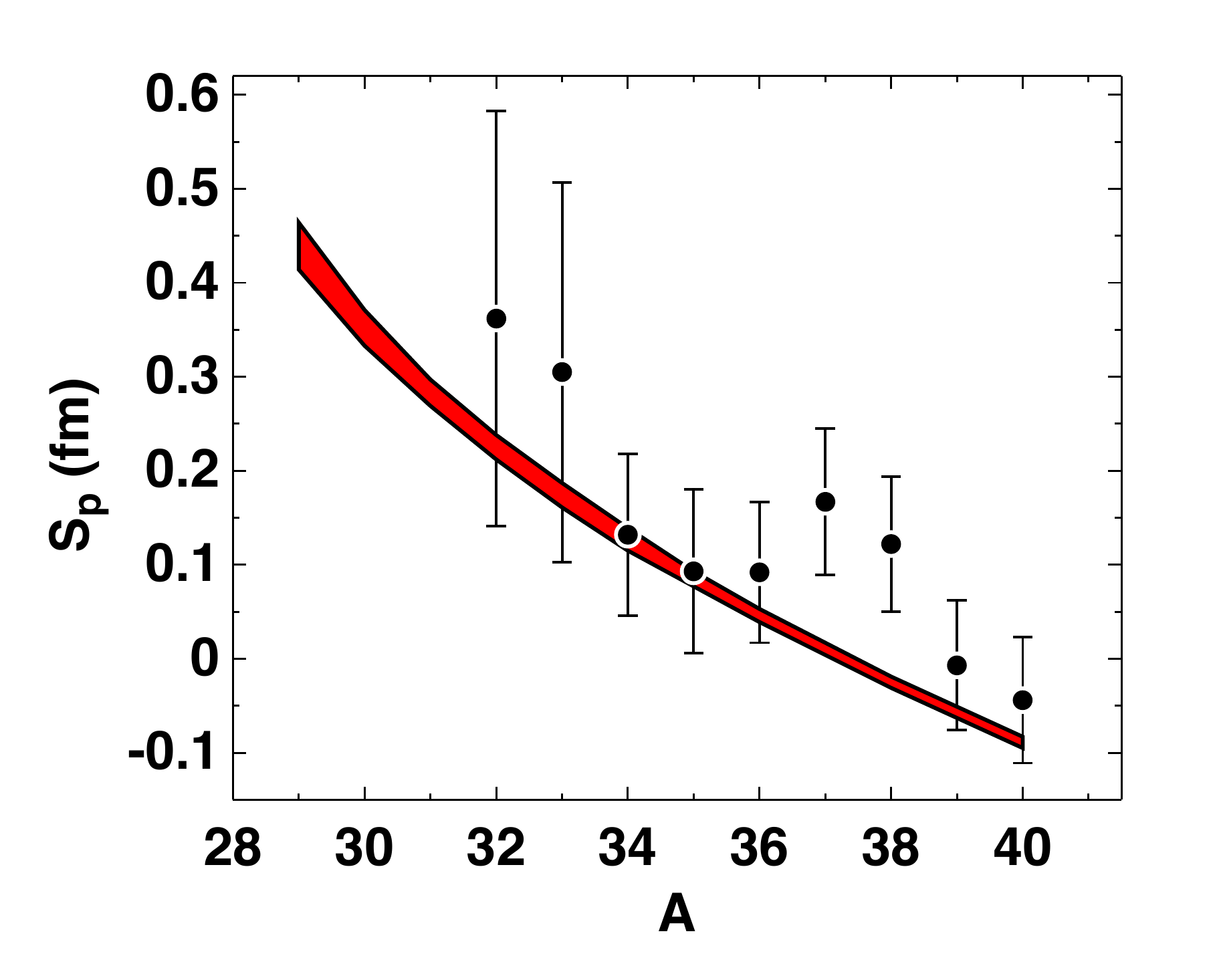}
\vspace*{0.1cm}
\caption{(Color online) Predicted proton skins of Argon isotopes 
as a function of the mass number, $A$.
The data points are from Ref.~\cite{Oz+02}.
} 
\label{ar}
\end{figure}

\section{Mirror nuclei} 
\label{mirr} 

\subsection{Symmetry of mirror nuclei} 
\label{sym} 

Assuming perfect charge symmetry, one has, in mirror nuclei,
\begin{equation}
R_n(Z,N) = R_p(N,Z) \; ,                     
\label{cs}
\end{equation}
a relation which we have verified to 
be exactly satisfied when Coulomb contributions and other charge-dependent effects are turned off.   
Applying the definition of the neutron skin, 
\begin{equation}
S_n(Z,N) = R_n(Z,N) - R_p(Z,N) \; ,                     
\label{skin1}
\end{equation}
we can then immediately conclude from Eq.~(\ref{cs}) that 
\begin{equation}
S_n (Z,N) = R_n(Z,N) - R_p(Z,N) = R_p(N,Z) - R_p(Z,N) \equiv \Delta R_{p} \; . 
\label{skin2} 
\end{equation}
Namely, the neutron skin of nucleus $(Z,N)$ would be {\it equal} to the difference between the proton radii
of the mirror pair in the presence of perfect charge symmetry. 
If charge radii could be measured accurately for mirror pairs in the desired mass range, then 
the neutron skin of the $(Z,N)$ nucleus 
could be obtained from Eq.~(\ref{skin2}) after theoretical considerations to account for 
 charge effects.                                                     
Thus, this could be an alternative, although perhaps equally challenging from the experimental side, to the 
anticipated parity-violating experiments~\cite{AB17}. 

\subsection{Radii and skins of mirror nuclei for $A\approx$50} 
\label{A50} 

\begin{table}                
\centering
\caption                                                    
{ Predicted proton skins, $S_p$, for the given $Z$ and $A$ and neutron skins, $S_n^{mirr}$, of the 
corresponding mirror nuclei.  
} 
\begin{tabular*}{\textwidth}{@{\extracolsep{\fill}}cccc}
\hline
\hline
Z & A & $S_p$ (fm) & $S_n^{mirr}$ (fm) \\ 
\hline     
10 & 16 & 0.422$\pm$0.022    & 0.333$\pm$0.016   \\ 
11 & 18 & 0.373$\pm$0.020   & 0.286$\pm$0.011   \\ 
17 & 31 & 0.180$\pm$0.012    & 0.091$\pm$0.006   \\ 
18 & 29 & 0.439$\pm$0.025    & 0.310$\pm$0.010   \\ 
\hline
\hline
\end{tabular*}
\label{tab2}
\end{table}

\begin{table}                
\centering
\caption                                                    
{ Proton skins, $S_p$, in the mass range 48-54.   
} 
\begin{tabular}{l l c }
\hline
\hline
Z \hspace{0.8in} & A \hspace{0.8in} & $S_p$ (fm) \\ 
\hline     
20 &48  & -0.181 $\pm$ 0.010  \\ 
28 &48  & 0.316 $\pm$ 0.021  \\ 
22 &50  & -0.112 $\pm$ 0.010  \\ 
28 &50  &  0.238 $\pm$ 0.016  \\ 
24 &52  & -0.048 $\pm$ 0.007  \\ 
28 &52  &  0.169 $\pm$ 0.013  \\ 
26 &54  &  0.008 $\pm$ 0.006  \\ 
28 &54  &  0.112 $\pm$ 0.013  \\ 
\hline
\hline
\end{tabular}
\label{tab3}
\end{table}

We now move to a specific range within medium mass nuclei, namely $A \approx 48-54$. This choice
can be motivated by the vicinity to $^{48}$Ca, whose neutron skin has already been and is likely to be in the future
the object of several investigations, both theoretical and experimental.
At the same time, the need to consider mirror pairs
limits the spectrum of realistic possibilities.

Table~\ref{tab4} displays the neutron skin of the neutron-rich isotones from Table~\ref{tab3} in relation to 
$\Delta R_p$ as defined in Eq.~(\ref{skin2}), with and without Coulomb effects. 
(Note that the latter case will not be addressed again and is shown here
only for numerical verification, since the two items appearing 
in parentheses in Table~\ref{tab4} are expected to be exactly equal to each 
other on grounds of elementary nuclear physics.) 

Increasing $\Delta R_p$ implies increasing the neutron skin, as one might reasonably expect unless
Coulomb effects were to reverse the relation in Eq.~(\ref{skin2}). 
Note, though, that quantitatively speaking Coulomb effects are significant. 

\begin{figure}[!t]
\vspace*{1cm}
\includegraphics[width=8.5cm]{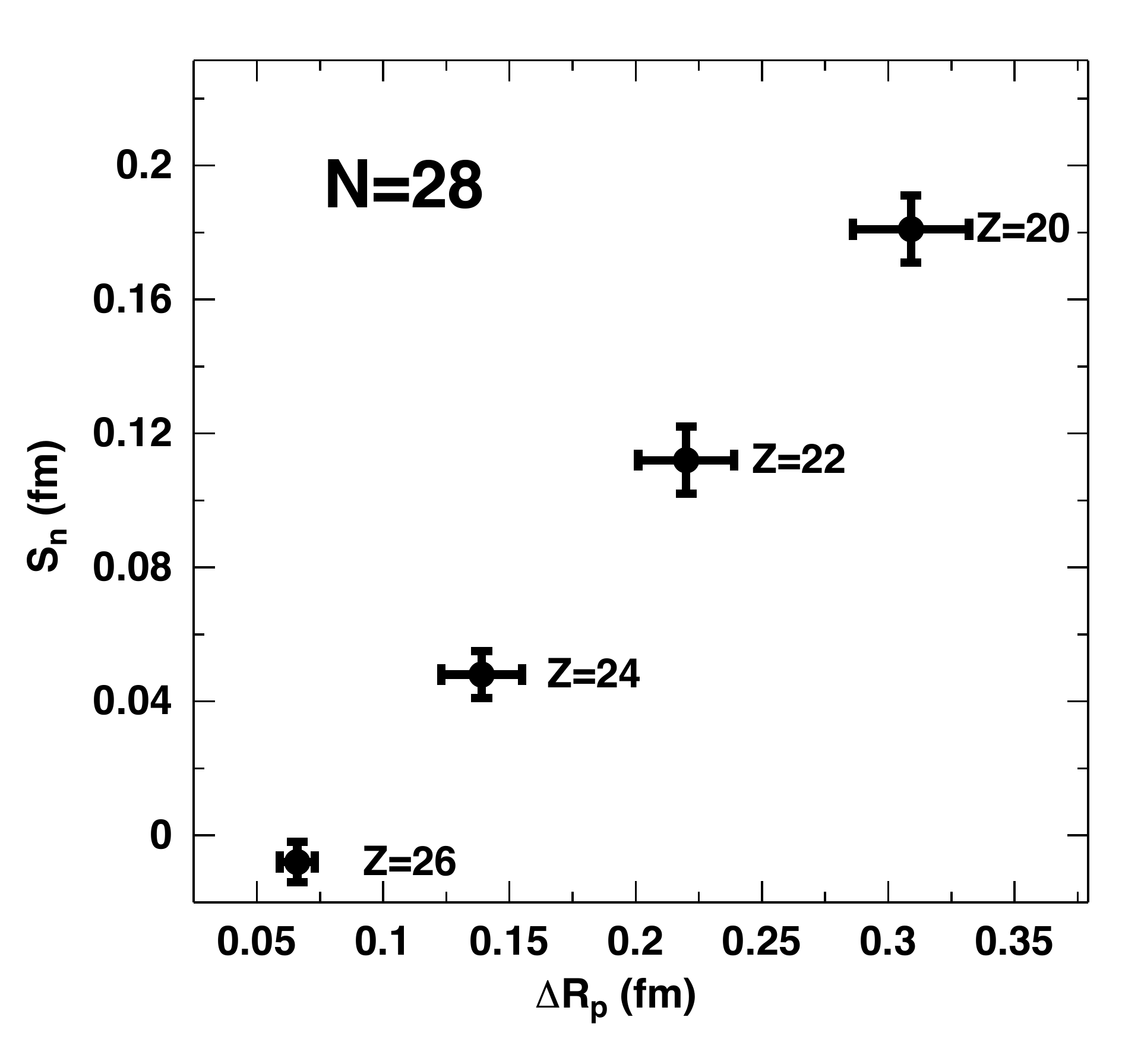}
\vspace*{0.1cm}
\caption{Graphical representation of Table~\ref{tab4}.}                                   
\label{t4}
\end{figure}

\begin{figure}[!t]
\vspace*{1cm}
\includegraphics[width=8.5cm]{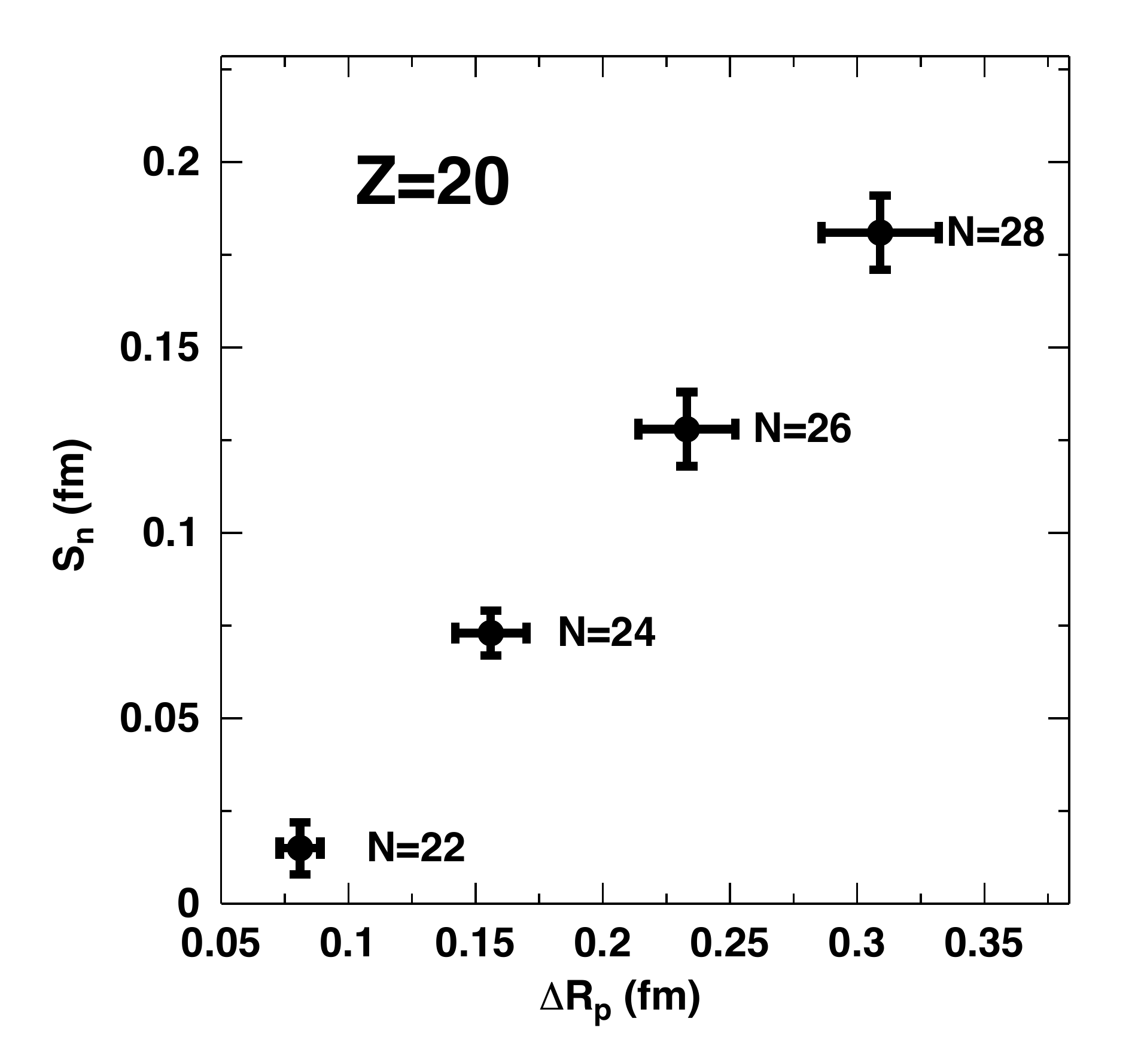}
\vspace*{0.1cm}
\caption{ Graphical representation of Table~\ref{tab5}.}                                   
\label{t5}
\end{figure}

\begin{figure}[!t]
\vspace*{1cm}
\includegraphics[width=8.5cm]{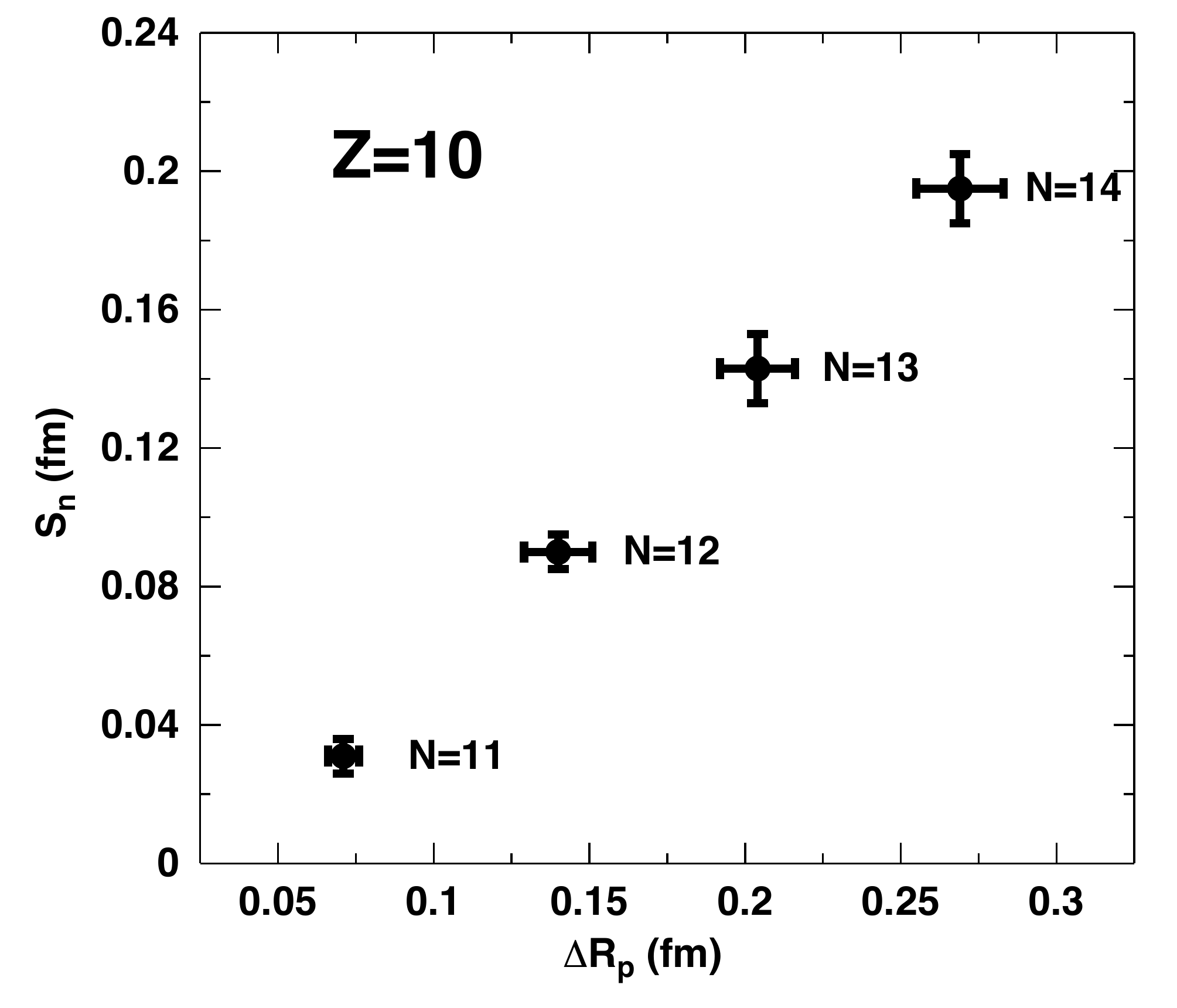}
\vspace*{0.1cm}
\caption{ Graphical representation of Table~\ref{tab6}.}                                   
\label{t6}
\end{figure}

Next we wish to explore the relation between $\Delta R_p$ and     
$S_{n}(Z,N)$ for other chains. In particular, we wish to investigate if and how such relation differs, 
quantitatively, among chains with different masses. For that purpose, 
we consider in Table~\ref{tab5} and \ref{tab6} two {\it isotopic} chains, one of them 
in a mass range considerably different than the one studied in Table~\ref{tab4}. 
A visual representation of Tables~\ref{tab4}, \ref{tab5}, and \ref{tab6} is provided in Figs.~\ref{t4}, \ref{t5},
 and \ref{t6}. 

The first observation is that,
for similar values of $\Delta R_p$, the corresponding values of
$S_{n}(Z,N)$ are approximately the same, regardless $Z$ and $N$.                                 
Also, in all three cases the relation is clearly linear.           
We stress again that the results shown in Figs.~\ref{t4},~\ref{t5}, and \ref{t6} are 
fundamentally distinct from the correlations discussed in Ref.~\cite{AB17}. The latter are obtained varying the parameters of Skyrme models (each model constrained to produce a chosen value of the neutron skin
in $^{208}$Pb) for a fixed mirror pair. Here, we explore to which extent our
microscopic EoS yields, within theoretical uncertainties, a unique relation between $S_n$ and $\Delta R_p$. 

The parameters of our predicted linear relation, 
\begin{equation}
S_n = a  (\Delta R_p) + b  \; ,                                   
\label{lin}
\end{equation}
based upon the three cases shown in 
Figs.~\ref{t4},~\ref{t5}, and \ref{t6}, can be summarized as 
\begin{equation}
a = 0.78 \pm 0.05 \; \; \; \; \; \; \; b = -0.044 \pm 0.016 \; .
\label{par}
\end{equation}
By means of Eqs.~(\ref{lin}-\ref{par}), 
a measurement of $\Delta R_p$ can then be promptly related to the neutron skin of the neutron-rich
nucleus in the mirror pair. 

\begin{table}                
\centering
\caption                                                    
{ Relation between the neutron skin of nucleus $(Z,N)$, $S_n(Z,N)$, and $\Delta R_p$ of the corresponding          
mirror pair
for the isotone chain $N$=28. 
The values in paranthesis are the results without Coulomb contribution (as a verification). 
} 
\begin{tabular*}{\textwidth}{@{\extracolsep{\fill}}cccc}
\hline
\hline
Z & N & $S_{n}(Z,N)$(fm) & $\Delta R_p$(fm) \\ 
\hline     
20 &28  & 0.181 $\pm$ 0.010 (0.229)  & 0.309 $\pm$ 0.023 (0.229)  \\ 
22 &28  & 0.112 $\pm$ 0.010 (0.162)  & 0.220 $\pm$ 0.019 (0.162)  \\ 
24 &28  & 0.048 $\pm$ 0.007 (0.103)  & 0.139 $\pm$ 0.016 (0.103)  \\ 
26 &28  & -0.008 $\pm$ 0.006 (0.049 )  & 0.066 $\pm$ 0.007 (0.049 ) \\ 
\hline
\hline
\end{tabular*}
\label{tab4}
\end{table}

\begin{table}                
\centering
\caption                                                    
{ Relation between the neutron skin of nucleus $(Z,N)$, $S_n(Z,N)$, and $\Delta R_p$                  
for the isotope chain $Z$=20. 
} 
\begin{tabular*}{\textwidth}{@{\extracolsep{\fill}}cccc}
\hline
\hline
Z & N & $S_{n}(Z,N)$(fm) & $\Delta R_p$(fm) \\ 
\hline     
20 &22  & 0.015 $\pm$ 0.007          & 0.081 $\pm$ 0.008          \\ 
20 &24  & 0.073 $\pm$ 0.006        & 0.156 $\pm$ 0.014          \\ 
20 &26  & 0.128 $\pm$ 0.010          & 0.233 $\pm$ 0.019          \\ 
20 &28  & 0.181 $\pm$ 0.010           & 0.309 $\pm$ 0.023          \\ 
\hline
\hline
\end{tabular*}
\label{tab5}
\end{table}

\begin{table}                
\centering
\caption                                                    
{ Relation between the neutron skin of nucleus $(Z,N)$, $S_n(Z,N)$,                                     
and $\Delta R_p$                  
for the isotope chain $Z$=10. 
} 
\begin{tabular*}{\textwidth}{@{\extracolsep{\fill}}cccc}
\hline
\hline
Z & N & $S_{n}(Z,N)$(fm) & $\Delta R_p$(fm) \\ 
\hline     
10 &11  & 0.031 $\pm$ 0.005          & 0.071 $\pm$ 0.005          \\ 
10 &12  & 0.090 $\pm$ 0.005        & 0.140 $\pm$ 0.011          \\ 
10 &13  & 0.143 $\pm$ 0.010          & 0.204 $\pm$ 0.012          \\ 
10 &14  & 0.195 $\pm$ 0.010           & 0.269 $\pm$ 0.014          \\ 
\hline
\hline
\end{tabular*}
\label{tab6}
\end{table}

Microscopic predictions do, of course, differ from one another.
Although EFT should, in principle, be a model-independent approach, even EFT-based predictions can differ between them,  depending, for instance, on the details of the 
input forces (e.g. cutoff) and the chosen many-body method. 
Moreover, the microscopically-predicted relations between two quantities or observables 
are not necessarily located on one of the Skyrme models correlations.
Here, we suggest that analyses such as the present one, combined with other microscopic predictions, 
are the best way to provide a global relation between the ``observables"
being studied (as well as their relation to the density dependence 
of the symmetry energy), accompanied by a meaningful theoretical uncertainty.

\section{Summary and Conclusions}                                                                  
\label{Concl} 

Microscopic predictions of the EoS for isospin-asymmetric nuclear matter have been applied to obtain proton and 
neutron skins of selected chains of nuclei. The calculations of the EoS are based on high-precision
chiral forces. 

First, we  presented proton skin predictions for a few isotopic
chains to observe some of their general fetures, particularly in 
comparison with neutron skins. We find that they are generally large, larger than neutron skins for comparable
values of proton-neutron asymmetry.                         
Our predictions compare well with available empirical information. 

We then moved the focus on to 
mirror nuclei in a specific mass range ($A \approx$ 48-54). At this point we took the opportunity to make some 
comments about and highlight differences with recent studies~\cite{AB17,YP} which have addressed those nuclei.

Using our microscopic predictions and their uncertainties, 
we constructed a correlation between the skin of a neutron-rich nucleus and the difference between the proton
radii of the corresponding mirror pair. We discussed the meaning and significance of such correlation in contrast 
to those characteristic of phenomenological studies. Given the {\it ab initio} nature of the EoS, we are in 
the position of exploring, for instance, the contribution of 3NF to the predictions, the impact of higher 
chiral orders,
and the order-by-order pattern of the chiral perturbation series. 

We conclude by highlighting the importance of taking into account microscopic predictions as a guide towards the planning of future measurements.

\section*{Acknowledgments}
This work was supported by 
the U.S. Department of Energy, Office of Science, Office of Basic Energy Sciences, under Award Number DE-FG02-03ER41270.

\end{document}